\documentclass[preprint,showpacs,preprintnumbers,amsmath,amssymb,nofootinbib]{revtex4}

\usepackage{dcolumn}% Align table columns on decimal point
\usepackage{bm}% bold math
\usepackage{braket}
\usepackage{amsmath}
\usepackage{txfonts}
\usepackage[pdftex]{graphicx,color}
\usepackage{float}
\begin{document}
\preprint{KUNS-2658 }

\title{Chiral primordial blue tensor spectra from the axion-gauge couplings}
\author{Ippei Obata}
\email{obata@tap.scphys.kyoto-u.ac.jp}

\affiliation{Department of Physics, Kyoto University, Kyoto, 606-8502, Japan}

\date{\today}% It is always \today, today,

\begin{abstract}
 We suggest the new feature of primordial gravitational waves sourced by the axion-gauge couplings, whose forms are motivated by the dimensional reduction of the form field in the string theory.
 In our inflationary model, as an inflaton we adopt two types of axion, dubbed the model-independent axion and the model-dependent axion, which couple with two gauge groups with different sign combination each other.
 Due to these forms both polarization modes of gauge fields are amplified and enhance both helicies of tensor modes during inflation.
 We point out the possibility that a primordial blue-tilted tensor power spectra with small chirality are provided by the combination of these axion-gauge couplings, intriguingly both amplitudes and chirality are potentially testable by future space-based gravitational wave interferometers such as DECIGO and BBO project.
 %We expect that the blue tensor spectra with little chirality becomes the new window to the axion phenomenology in the early universe.
\end{abstract}

\pacs{Valid PACS appear here}% PACS, the Physics and Astronomy

\maketitle

\tableofcontents

\section{Introduction}

 The primordial inflation quantum-mechanically accounts for the origin of Cosmic Microwave Background radiation (CMB) temperature anisotropy and large scale structure in our present universe.
 Through these observations, it allows us to explore the high energy physics by means of inflation.
 One of the most important prediction is that it also provides vacuum fluctuations of gravitons, called primordial gravitational waves (GWs), whose amplitude is directly related to the energy scale of inflation in the standard scenario.
 We can search for the imprint of primordial gravitational waves on the polarization of CMB B-mode, conventionally parameterized by the tensor-to-scalar ratio ~$r$ .
 Unfortunately, however, we have not yet detected this signal and recent BICEP2/Keck Array collaboration has reported the following upper bound on its amplitude ~$r \lesssim 0.07$ \cite{Array:2015xqh}.
 Moreover, due to the decreasing of Hubble parameter during inflation, its scale dependence is always red.
 Therefore, it is also getting harder to detect it in future gravitational wave experiments.
 In fact, for the practically testable GW amplitude the space-based interferometers such as DECIGO (not Ultimate DECIGO) and BBO project require ~$r = \mathcal{O}(0.1)$ ~in the case of a standard inflationary model \cite{Kawamura:2011zz, Crowder:2005nr}.

 From the theoretical point of view, however, there is a room to suppose another source of metric tensor modes based on the fundamental theory.
 In superstring theory, there is a plentitude of pseudo-scalar fields, called string axions, which arise from the compactification of the extra-dimensional manifold \cite{Svrcek:2006yi, Arvanitaki:2009fg}.
 It is known that the string axion is one of the best motivated candidate of the inflaton due to its shift symmetry at perturbative level, which protects the flatness of the potential from the radiative correction of other coupled matters \cite{Freese:1990rb, Kim:2004rp}.
 %Due to this concept, it is worth probing the quantitatively new phenomena through the primordial gravitational waves from the axionic inflation.
 Intriguingly, the rolling axion amplifies one helicity mode of the gauge field via its Chern-Simons interaction, so that the enhanced gauge field could provide the parity-violating tensor spectrum, in both cases of the Abelian gauge field and the non-Abelian gauge field \cite{Sorbo:2011rz, Barnaby:2011vw, Cook:2011hg, Barnaby:2011qe, Anber:2012du, Barnaby:2012xt, Linde:2012bt, Mukohyama:2014gba, Ferreira:2014zia, Namba:2015gja, Ferreira:2015omg, Domcke:2016bkh, Guzzetti:2016mkm, Peloso:2016gqs, Garcia-Bellido:2016dkw, Dimastrogiovanni:2012ew, Adshead:2013qp, Obata:2014loa, Maleknejad:2016qjz, Dimastrogiovanni:2016fuu}.
 Furthermore, due to the increasing of the velocity of axion, its amplitude is naturally blue-tilted and becomes sizable higher than frequency range of CMB scales.
 The main motivation of this study is to search for the axion phenomenology through chiral primordial gravitational waves sourced by the axion-gauge coupling, which is potentially testable by future space-based gravitational wave experiments or pulsar timing arrays.

 In recent studies \cite{Domcke:2016bkh, Garcia-Bellido:2016dkw}, it was shown that for an observable GW amplitude a preferable range of axion decay constant is around GUT scales, remarkably which is naturally predicted by the low energy supergravity of the string theory \cite{Svrcek:2006yi}.
% Based on this fact, in this paper we explore the specific form of axion-gauge couplings derived from the dimensional reduction of the heterotic ~$\text{E}_8 \times \text{E}_8$ ~superstring and reconsider the generation of tensor modes.
 Motivated by this result, in this paper we explore the novel feature of blue tensor spectrum produced by the string compactifications.
 In the context of the dimensional reduction of string theory, two types of axion-like particle generally appear as the zero mode of Kalza-Klein expansion of the form field, traditionally called the ``model-independent axion" and the ``model-dependent axion".
 Both of them couple with the gauge field through the anomaly cancellation mechanism in ten-dimentional theory.
 However, the way of coupling to gauge fields is different between these axions: the model-independent axion acquires the gauge interaction from the kinetic term of the form field, while that of the model-dependent axion derives from the one-loop Green-Schwarz counter terms of gauge anomalies.
 As a result, the sign combination of the coupled Chern-Simons terms are generally different between these axions.
 Therefore, the opposite helicities of gauge fields each other could play an important role if these couplings are effective during inflation.
 %After truncation to four dimension, they generically couple with the gauge field due to the chiral anomaly cancellation mechanism in ten-dimensional theory. 
 %In order to break the large gauge symmetry down to subgroups
 %Interestingly, the sign of combination of gauge interactions to each axion is not necessarily the same.
 %Namely, it is not necessarily true that only one helicity mode of the gauge field experiences a tachyonic instability during inflation.
 %Inspired by this property, in this work we suggests the new feature of providing primordial blue tensor spectra (not ``spectrum"!) from the combination of these two axion-gauge couplings.

 Inspired by this property, in this work we suggests the possibility of generating blue tensor spectra (not ``spectrum"!) sourced by two gauge groups coupled to these two types of axion.
 In the conventional scenario, in order to get a sizable tensor spectrum in small scales its parity must be almost totally violated because only one polarization mode of gauge field is amplified during inflation.
 We find that, however, if these two axions play a role of inflaton, the different polarization modes of each gauge group coupled to two axions grow up and could provide totally parity-conserved or slightly parity-violated blue tensor spectra, both of which add completely new features to the axion phenomenology in early universe.
 Remarkably whose amplitudes and chiralities are potentially testable with future space-based interferometer DECIGO or BBO project.
 %Especially, in the latter case we expect that such a small but non-negligible parity-violation feature will be a smoking-gun of the axion phenomenology in early universe.

 %It is known that the string axion is one of the best motivated candidate of inflaton due to its shift symmetry which protects the flatness of potential from radiative corrections.
 This paper is organized as follows.
 In section \ref{setup}, we derive two types of axion and their form of gauge couplings from the dimensional reduction of the form field.
 In section \ref{model}, we build the concrete inflationary models where two types of axion occur inflation coupled to different sign combination of Abelian gauge sectors, which are generally included in the original two gauge groups.
 Then in this section we describe the time evolution of tensor power spectra numerically.
 Simultaneously, we also check the production of the inflaton perturbation sourced by gauge fields and show that the power spectrum of curvature perturbation is consistent with the non-detection of primordial black holes (PBHs) during inflation.
 The final section is devoted to conclusion and discussion.
 Note that we set ~$\hbar = c = 1$ ~in our units.

\section{The axion-gauge coupling from the string theory \label{setup}}

 In this section, we briefly derive two types of the axion-gauge interaction from the string theory, especially the weakly coupled heterotic $\text{E}_8\times\text{E}_8$ superstring as an example \cite{Choi:1985je, Ibanez:1986xy, Blumenhagen:2005ga}\footnote{The result of this section is also derived from the strong coupling limit of the $\text{E}_8\times\text{E}_8$ heterotic superstring (M-theory) \cite{Banks:1996ss, Choi:1997an}.}.
 %Suppose that our world is a ten-dimensional space-time of the form ~$\text{M}_4 \times \text{M}_6$ : $\text{M}_4$ ~is a four-dimensional Lorentz space-time and ~$\text{M}_6$ ~is a compact six-dimensional manifold.
 The relevant bosonic part of its low-energy effective action in the string frame reads \cite{Polchinski}
\begin{equation}
S = \dfrac{1}{2\kappa_{10}^2}\int_{\text{M}_{10}}e^{-2\phi_{10}}\left[ d^{10}x\sqrt{-g_{10}}R - \dfrac{1}{2}H_3\wedge*_{10}H_3 - \dfrac{\kappa_{10}^2}{g_s^2}\text{tr}(F_2\wedge*_{10}F_2) \right] \ ,
\end{equation}
where ~$R$ ~is Ricci scalar (~$g_{n}$ is a determinant of a ~$n$-dimensional metric), ~$H_3$ ~is the field strength of the Neveu-Schwarz 2-form ~$B_2$ , and ~$F_2$ ~is the field strength of ~$\text{E}_8\times\text{E}_8$ ~gauge group living in the ten-dimensional space-time $\text{M}_{10}$.
 This action contains three coupling constants: the dilaton coupling ~$e^{\phi_{10}}$ , the Yang-Mills coupling ~$g_s$ ~and the gravitational coupling ~$\kappa_{10}$ ~with dimensions of (length)$^0$, (length)$^3$ and (length)$^4$ respectively.
 We write the ~``$\text{tr}$" ~for the trace in the fundamental representation of ~$\text{E}_8$ .
 The symbol ``$*$" is the Hodge dual operator so that ~$X\wedge*_{n}X = d^{n}x\sqrt{|g_n|}X_{M_1M_2...M_p}X^{M_1M_2...M_p}/p!$ ~is satisfied if ~$X$ ~is a p-form.
 Hereafter, we usually write the products and powers of forms without the wedge symbol ~``$\wedge$" ~to keep notations compact, so that ~$X^2 \equiv X \wedge X$ ~and ~$|X|_n^2 \equiv X \wedge *_n X$ .
 Note that we assume that the dilaton has been stabilized and integrated out in the action.

 The gauge-invariant field strength ~$H$ ~is given by
~$
H_3 = dB_2 - \frac{\alpha'}{4}(\omega_{3Y} - \omega_{3L}) \ ,
$~
where ~$\alpha'$ ~is the string tension and ~$\omega_{3Y} \ , \omega_{3L}$ ~are the Yang-Mills and Lorentz Chern-Simons 3-forms ensuring absence of anomalies, so that the following Bianchi identity
\begin{equation}
0 = \int_\Sigma dH_3 = -\frac{\alpha'}{4}\int_\Sigma\left[\text{tr}(F_2^2) - \bm{tr}(R_2^2) \right] \label{eq: Bianchi}
\end{equation}
holds for the integration on any four-dimensional submanifold ~$\Sigma$ ~embedded in the internal manifold.
 Note that the Riemann tensor is regarded as a 2-form ~$R_2=(R_2){^p}_q = d\omega{_1}{^p}_q + \omega{_1}{^p}_r \wedge \omega{_1}{^r}_q$ ~which is a tangent space matrix represented by the connection 1-form ~$\omega{_1}{^p}_q$ ~and the ``$\bm{tr}$" ~is over the trace on the tangent space indices.
 As to the trace of the gauge field, we explicitly take care of two ~$\text{E}_8$ ~gauge factors ~$F^{(1)}_2$ ~and ~$F^{(2)}_2$ ~and separate it into ~$\text{tr}(F_2^2) = \text{tr}(F{^{(1)2}_2}) + \text{tr}(F{^{(2)2}_2})$ .

 Let us derive the axion-gauge couplings from the above action.
 As to the metric, we assume that it is separated into the four-dimensional space-time ~$\text{X}_4$ ~and the conpactified six-dimensional space ~$\text{Y}_6$ : $ds^2(\text{M}_{10}) = ds^2(\text{X}_4) + ds^2(\text{Y}_6)$ .
 Pseudo-scalar fields with axion-like properties generically arise from the component of~$B_2$ ~field.
 Focusing on the zero-mode, we expand it as
\begin{equation}
B_2(\text{M}_{10}) = b_2(\text{X}_4) + l_s^2\sum_{k=1}^{n}b_{0(k)}(\text{X}_4)~\omega_k(\text{Y}_6) \ ,
\end{equation}
where ~$b_2$ ~and ~$b_{0(k)}$ ~are the 2-form and the 0-form.
 As we show later, ~$b_2$ ~is reduced to the model-independent axion; namely it does not depend the choice of ~$\text{Y}_6$ ~very much.
 On the other hand ~$b_{0(k)}$ ~depends on the compactification of ~$\text{Y}_6$ , so it is called the model-dependent axion.
 Here, ~$n = \text{dim}~H^2(\text{Y}_6)$ ~is the dimension of the basis of closed non-exact 2-form ~$\omega_k$ ~dual to the 2-cycles ~$C_l$ ~on ~$\text{Y}_6$ ~representing a basis of ~$H_2(\text{Y}_6)$ ~modulo torsion: $\int_{C_l}\omega_k = \delta_{lk}$ .
 For dimensional reasons, we introduced the string length scale ~$l_s=2\pi\sqrt{\alpha'}$ .
% Defining ~$*_4db_2 \equiv e^{2\phi_{10}}db_0$ ~and ~$*_4db_{0(k)} \equiv e^{2\phi_{10}}db_{2(k)}$ , we can get
%\begin{equation}
%*_{10}dB_2 = \left[ db_0\text{vol}_6 + \sum_{k=1}^{b_2(Y)}db_{2(k)}\hat{\omega}_k \right] \ ,
%\end{equation}
%where ~$\text{vol}_6$ ~is the normalized 6-form on ~$\text{M}_6$ ~so that ~$\int_{\text{M}_6}\text{vol}_6 = 1$ ~holds.
 By the dimensional reduction from the kinetic term of ~$B_2$ ~field, we can simply get the following four-dimensional action
%\begin{equation}
%\int d^4x\sqrt{-g}\left[ -\dfrac{1}{2}\partial_\mu b_0 \partial^\mu b_0  - \sum_{k=1}\dfrac{1}{2}\partial_\mu b_{0(k)} \partial^\mu b_{0(k)} - \dfrac{1}{4}b_0\left(\text{tr}(F_{1 \mu\nu}\tilde{F}_{1}^{\mu\nu}) + \text{tr}(F_{2\mu\nu}\tilde{F}_2^{\mu\nu})\right) \right]
%\end{equation}
%\begin{equation}
%\dfrac{e^{-2\phi_{10}}}{4\kappa_{10}^2}V_6\int_{\text{M}_4}\left[ -\dfrac{1}{2}|h_3|_4^2 - \sum_{i j}Y^{ij}db_{0(i)}\wedge*_4db_{0(j)} + \dfrac{\theta}{\pi}\left\{dh_3 + \dfrac{\alpha'}{4}(\text{tr}F^2 - \text{tr}R^2)\right\} \right] \ ,
%\end{equation}
\begin{equation}
\dfrac{e^{-2\phi_{10}}}{4\kappa_{10}^2}L^6\int_{\text{M}_4}\left[ -\dfrac{1}{2}|h_3|_4^2 - \left(\dfrac{l_s}{L}\right)^4\sum_{k=1}^n|db_{0(i)}|_4^2 + \dfrac{\theta}{\pi}\left\{dh_3 + \dfrac{\alpha'}{4}(\text{tr}(F_2^2) - \bm{tr}(R_2^2))\right\} \right] \ ,
\end{equation}
where ~$L$ ~is the unit length of ~$\text{Y}_6$ , $h_3$ ~is the four-dimensional component of ~$H_3$ ~and %~$Y_{ij} \equiv l_s^4/V_6\int_{\text{Y}_6} \omega_i \wedge *_6\omega_j$
we chose the basis ~$\omega_i$ ~to satisfy ~$\int_{\text{Y}_6} \omega_i \wedge *_6\omega_j = L^2\delta_{ij}$ .
 Here, we introduced a Lagurange multiplier ~$\theta$ ~in order to satisfy the Bianchi identity in our four-dimensional space-time.
 Integrating ~$h_3$ ~out and canonically normalizing each variable, we get the kinetic terms of axions and Chern-Simons terms coupled to the model-independent axion
\begin{equation}
\int d^4x\sqrt{-g_4}\left[ -\dfrac{1}{2}\partial_\mu a \partial^\mu a  - \sum_{k=1}\dfrac{1}{2}\partial_\mu a_{k} \partial^\mu a_{k} + \dfrac{\lambda}{4}\dfrac{a}{f_\theta}\left(\text{tr}(F^{(1)}_{\mu\nu}\tilde{F}_{(1)}^{\mu\nu}) + \text{tr}(F^{(2)}_{\mu\nu}\tilde{F}_{(2)}^{\mu\nu})\right) \right] \label{eq: moin} \ ,
\end{equation}
where
\begin{equation}
f_\theta \equiv \dfrac{e^{-\phi_{10}}L^3}{2\sqrt{2}\pi\kappa_{10}} = \dfrac{M_p}{2\sqrt{2}\pi} \ , \qquad \lambda \equiv 2\pi\alpha' f_\theta^2 = \dfrac{1}{16\pi^3}\left(\dfrac{M_p}{M_s}\right)^2
\end{equation}
is a decay constant of model-independent axion and a dimensionless coupling constant of its Chern-Simons gauge interaction characterized by the string mass scale ~$M_s \equiv l_s^{-1}$ .

 On the other hand, as to the model-dependent axions, they acquire the gauge couplings from the Green-Schwarz anomaly cancellation action \cite{Green:1984sg}
\begin{equation}
S_{\text{GS}} = \dfrac{1}{288(2\pi)^5\alpha'}\int_{\text{M}_{10}} B\left[ \text{Tr}(F_2^4) - \dfrac{1}{300}\left[\text{Tr}(F_2^2)\right]^2 - \dfrac{1}{10}\text{Tr}(F_2^2)\bm{tr}(R_2^2) + ... \right] \label{eq: GS} \ ,
\end{equation}
where the trace ~``$\text{Tr}$" ~is over the adjoint representation of ~$E_8$, which is related to the $\text{tr}$ as ~$\text{Tr} = 30~\text{tr}$.
 Note that we have omitted the purely gravitational anomaly cancellation terms ``..." in the above action in order to focus on the coupling of the gauge field.
 In terms of the following relationships ~$\text{Tr}(F_2^4) = [\text{Tr}(F_2^2)]^2/100, \ \text{Tr} = 30~\text{tr}$, we can rewrite \eqref{eq: GS} into
\begin{equation}
\dfrac{1}{48(2\pi)^3}\sum_{k=1}^n\int_{\text{X}_4}b_{0(k)} ~\text{tr}(F_2^2) \int_{\text{Y}_6} \omega_k\left[ \text{tr}(F_2^2) - \dfrac{1}{2}\bm{tr}(R_2^2) + ... \right] \ . \label{eq: GS2}
\end{equation}
 Then separating ~$\text{tr}(F_2^2) = \text{tr}(F{^{(1)2}_2}) + \text{tr}(F{^{(2)2}_2})$ ~and using the Bianchi identity \eqref{eq: Bianchi},
%\begin{equation}
%\int_{\text{Y}_6} \omega_k\left[ \text{tr}(F_1^2) + \text{tr}(F_2^2) - \bm{tr}(R^2) \right] = -\int_{\text{Y}_6} \omega_k ~dH = 0 \ ,
%\end{equation}
we get the following interactions of axion-gauge couplings
\begin{equation}
\eqref{eq: GS2} \supset \dfrac{1}{48(2\pi)^3}\sum_{k=1}^n I_k \int_{\text{X}_4}b_{0(k)} ~\left( \text{tr}(F_2^{(1)2}) - \text{tr}(F_2^{(2)2}) \right) \ , \qquad I_k \equiv \int_{\text{Y}_6} \omega_k\left[ \text{tr}(F_2^{(1)2}) - \dfrac{1}{2}\bm{tr}(R_2^2) \right] \ .
\end{equation}
 Therefore, the coupling of model-dependent axions to gauge fields in the four dimension has the following form
\begin{align}
\sum_{k=1}^n\int d^4x\sqrt{-g_4}\dfrac{\lambda_k}{4}\dfrac{a_k}{f_k}\left( \text{tr}(F^{(1)}_{\mu\nu}\tilde{F}_{(1)}^{\mu\nu}) - \text{tr}(F^{(2)}_{\mu\nu}\tilde{F}_{(2)}^{\mu\nu}) \right) \ , \label{eq: mode} 
\end{align}
where
\begin{equation}
f_k \equiv \left(\dfrac{l_s}{L}\right)^2f_\theta \ , \qquad \lambda_k \equiv \dfrac{I_k}{12(2\pi)^3} \ .
\end{equation}
 Remarkably, the sign of each Chern-Simons interaction is opposite.

 Thus, we get the two types of axion-gauge couplings in \eqref{eq: moin} and \eqref{eq: mode} from the dimensional reduction of ~$\text{E}_8 \times \text{E}_8$ ~superstring.
 Other than the heterotic string theory, these sign combination of axion-gauge couplings could also be derived from the compactifications with D-branes through the generalized Green-Shwarz mechanism \cite{Blumenhagen:2006ci}.
 Below this section, we show that this interaction forms could provide the blue tensor spectra whose both polarization modes are largely enhanced in small scales.

\section{Axionic inflationary models with gauge fields \label{model}}

 In this section, we present axionic inflationary models with two gauge groups inspired by the dimensionally reduced string theory.
 In this work, however, we focus on the dynamics of Abelian sector ~$F^{(i)}_{\mu\nu} = \partial_\mu A^{(i)}_{\nu} - \partial_\nu A^{(i)}_{\mu} \ (i = 1, 2, ...)$ ~for an analytical reason and do not lead to something specific gauge groups from string compactifications.
 %The detail construction of our inflationary model from the perspective of flux compactifications in string theory is beyond the scope of this paper and we leave such an analysis to future work.
 The detail string theory embeddings of our scenarios and their phenomenological features are beyond the scope of this paper and we leave such an analysis to future work.
 As to the detail calculation about the particle production of gauge field and the amplification of tensor perturbations according to it, we discuss these issues in our appendix.

\subsection{Inflation with model-dependent axion-gauge couplings}

 Firstly, we consider the case where a model-dependent axion ~$\varphi$ ~coupled with two gauge groups occurs the inflation.
 For simplicity, we neglect the higher interaction of the non-Abelian gauge sector and we treat Abelian gauge fields embedded in the original gauge groups.
 So we set up the following action
\begin{align}
\int d^4x\sqrt{-g_4}&\left[ \dfrac{M_p^2}{2}R
- \dfrac{1}{2}(\partial \varphi)^2 - V(\varphi) - \sum_{n=1}^{\mathcal{N}}\dfrac{1}{4}F^{(n)}_{\mu\nu}F_{(n)}^{\mu\nu} - \sum_{m=1}^{\mathcal{M}}\dfrac{1}{4}F^{(m)}_{\mu\nu}F_{(m)}^{\mu\nu} \right. \notag \\
&\left. + \dfrac{1}{4}\dfrac{\varphi}{f}\left( \sum_{n=1}^{\mathcal{N}}F^{(n)}_{\mu\nu}\tilde{F}_{(n)}^{\mu\nu} - \sum_{m=1}^{\mathcal{M}}F^{(m)}_{\mu\nu}\tilde{F}_{(m)}^{\mu\nu} \right) \right] \ .
\end{align}
As to the metric, we assume a flat FLRW metric ~$ds^2 = -dt^2 + a(t)^2d\bm{x}^2 = a(\tau)^2[-d\tau^2 + d\bm{x}^2]$.
 The gauge field ~$A^{(i)}(\tau, \bm{x})$ ~interacts with the homogeneous axion field ~$\bar{\varphi}(t)$ ~via their Chern-Simons terms.
 In the following gauge conditions ~$A^{(i)}_0 = \partial_j A^{(i)}_j = 0$ , the equations of motion for each gauge field are given by
\begin{align}
\partial_\tau^2A^{(n)}_i - \nabla^2A^{(n)}_i - \dfrac{\bar{\varphi}'}{f}\epsilon_{ijk}\partial_jA^{(n)}_k &= 0 \ , \\
\partial_\tau^2A^{(m)}_i - \nabla^2A^{(m)}_i + \dfrac{\bar{\varphi}'}{f}\epsilon_{ijk}\partial_jA^{(m)}_k &= 0
\end{align}
(the prime means the derivative with respect to ~$\tau$ ).
 In order to describe the generation of gauge field induced by the rolling axion, we quantize ~$A^{(i)}_i$ ~in the basis of the circular polarization vectors ~$e^\pm_i(\hat{\bm{k}})$ , where the helicity operators ~$\hat{A}^{(i)}_\pm(\bm{k}, \tau)$ ~decompose into annihilation and creation operators ~$\hat{A}^{(i)}_{\pm}(\bm{k}, \tau)= a{^{(i)}}_{\pm}(\bm{k})A^{(i)}_{\pm}(k, \tau) + a^{(i)}{^{\dagger}}_{\pm}(-\bm{k})A^{(i)*}_{\pm}(k, \tau) $, obeying 
~$
[a{^{(i)}}_{\lambda\bm{k}} , a{^{(j)\dagger}}_{\lambda'-\bm{k}'}] = \delta_{ij}\delta_{\lambda\lambda'}(2\pi)^3\delta^3(\bm{k} + \bm{k}')
$.
 Hence the equations of motion for the mode functions ~$A^{(i)}_\pm(k, \tau)$ ~read
\begin{align}
\left[ \partial_{\tau}^2 + k^2 \pm\dfrac{2k\xi}{\tau}\right]A^{(n)}_{\pm} = 0 \ , \\ 
\left[ \partial_{\tau}^2 + k^2 \mp\dfrac{2k\xi}{\tau}\right]A^{(m)}_{\pm} = 0 \ ,
\end{align}
where ~$\xi \equiv \dot{\bar{\varphi}}/(2fH)$ ~(the dot means the derivative with respect to ~$t$ ) relates the velocity of the axion and hereafter we treat it as a positive value.
 We can see that ~$A^{(n)}_+$ ~and ~$A^{(m)}_-$ ~experience a tachyonic instability in the same time interval ~$|k\tau| \lesssim 2\xi$ ~and grow up by a factor ~$e^{\pi\xi}$ (see \eqref{eq: A}, \eqref{eq: A'}), especially the enhanced polarization mode is opposite.
 Hence, as we show later in total both polarization modes of the gauge fields from each gauge group could contribute the generation of gravitational waves.
 On the other hand, ~$A^{(n)}_-$ ~and ~$A^{(m)}_+$ ~are not produced during inflation and can therefore be ignored in the amplification of perturbations. 

 The transverse traceless tensor modes ~$h_{ij}(\tau, \bm{x})$ ~are given by the spatial component of the metric ~$g_{ij} = a^2(\delta_{ij} + h_{ij})$.
 In terms of the canonically normalized variable ~$\psi_{ij} = aM_ph_{ij}/2$, from the Einstein equations we get the equation of motion for tensor modes
\begin{equation}
\left[\partial_\tau^2 - \nabla^2 - \dfrac{2}{\tau^2} \right]\psi_{ij} = -\dfrac{a^3}{M_p}\sum_{i=1}^{\mathcal{N} + \mathcal{M}}(E^{(i)}_iE^{(i)}_j + B^{(i)}_iB^{(i)}_j)^{TT} \ ,
\end{equation}
where ~$E^{(i)}_i, \ B^{(i)}_i$ ~are the electric and magnetic component of the gauge field and ~$TT$ ~denotes the transverse traceless projection of the spatial component of energy-momentum tensor of gauge field.
 The two helicities of tensor modes ~$\hat{\psi}_\pm(\bm{k})$ ~are obtained by multiplying ~$\psi_{ij}$ ~by the circular polarization tensors ~$e^{ij}_\mp(\hat{\bm{k}}) = e^{i}_\mp(\hat{\bm{k}})e^{j}_\mp(\hat{\bm{k}})$ .
 Ignoring the contribution of ~$\hat{A}^{(n)}_-$ ~and ~$\hat{A}^{(m)}_+$, one can then find
\begin{align}
\left[\partial_\tau^2 + k^2 - \dfrac{2}{\tau^2} \right]&\hat{\psi}_\pm(\bm{k}) \simeq -\dfrac{1}{aM_p}\int\dfrac{d\bm{p}}{(2\pi)^3} e^{i}_\mp(\hat{\bm{k}})e^{j}_\mp(\hat{\bm{k}}) \notag \\
&\times\left[e^+_i(\hat{\bm{p}})e^+_j(\hat{\bm{k}-\bm{p}})\sum_{n=1}^\mathcal{N}\hat{J}^{(n)}_+(\bm{p}, \bm{k}-\bm{p}) + e^-_i(\hat{\bm{p}})e^-_j(\hat{\bm{k}-\bm{p}})\sum_{m=1}^\mathcal{M}\hat{J}^{(m)}_-(\bm{p}, \bm{k}-\bm{p}) \right] \label{eq: tensor} \ ,
\end{align}
where
\begin{equation}
\hat{J}^{(i)}_\pm(\bm{p}, \bm{k}-\bm{p}, \tau) \equiv \hat{A}^{(i)}_\pm{'}(\bm{p}, \tau)\hat{A}^{(i)}_\pm{'}(\bm{k} - \bm{p}, \tau) + p|\bm{k} - \bm{p}|\hat{A}^{(i)}_\pm(\bm{p}, \tau)\hat{A}^{(i)}_\pm(\bm{k} - \bm{p}, \tau) \ .
\end{equation}
 Therefore, the tensor modes composite of two solutions ~$\hat{\psi}_\pm = \hat{\psi}_{V} + \hat{\psi}_{S\pm}$ : ~$\hat{\psi}_V$ ~is the homogeneous solution which corresponds to the vacuum fluctuations and ~$\hat{\psi}_{S\pm}$ is the peculiar solution sourced by two gauge quanta ~$\hat{A}^{(n)}_+ \ , \hat{A}^{(m)}_-$.
%\begin{align}
%\hat{\psi}_{S+}(k, \tau) &\simeq -\dfrac{1}{M_p}\int_{-\infty}^{0} d\tau_1\dfrac{G_k(\tau, \tau_1)}{a(\tau_1)}\int\dfrac{d\bm{p}}{(2\pi)^3}e^{i}_-(\hat{\bm{k}})e{^+}_i(\hat{\bm{p}})e^{j}_-(\hat{\bm{k}})e{^+}_j(\hat{\bm{k}-\bm{p}}) \notag \\
%&\times \left[ \hat{A}^{(1)}_+{'}(\bm{p})\hat{A}^{(1)}_+{'}(\bm{k}-\bm{p}) + p|\bm{k} - \bm{p}|\hat{A}^{(1)}_+(\bm{p})\hat{A}^{(1)}_+(\bm{k}-\bm{p}) \right] \ , \\
%\hat{\psi}_{S-}(k, \tau) &\simeq -\dfrac{1}{M_p}\int_{-\infty}^{0} d\tau_1\dfrac{G_k(\tau, \tau_1)}{a(\tau_1)}\int\dfrac{d\bm{p}}{(2\pi)^3}e^{i}_+(\hat{\bm{k}})e{^-}_i(\hat{\bm{p}})e^{j}_+(\hat{\bm{k}})e{^-}_j(\hat{\bm{k}-\bm{p}}) \notag \\
%&\times \left[ \hat{A}^{(2)}_-{'}(\bm{p})\hat{A}^{(2)}_-{'}(\bm{k}-\bm{p}) + p|\bm{k} - \bm{p}|\hat{A}^{(2)}_-(\bm{p})\hat{A}^{(2)}_-(\bm{k}-\bm{p}) \right] \ ,
%\end{align}

 Let us calculate the two-point correlation function of the graviton ~$\hat{h}_\pm = 2\hat{\psi}_\pm/(aM_p) = \hat{h}_{V} + \hat{h}_{S\pm}$.
 Since ~$\hat{h}_{V} \ , \hat{h}_{S\pm}$ ~are statistically independent, the correlation function represents
~$
\langle\hat{h}_{\pm}(\bm{k})\hat{h}_{\pm}(\bm{k}')\rangle = \langle\hat{h}_{V\pm}(\bm{k})\hat{h}_{V\pm}(\bm{k}')\rangle + \langle\hat{h}_{S\pm}(\bm{k})\hat{h}_{S\pm}(\bm{k}')\rangle
$.
 We are interested in the correlation function of the source term.
 In view of the conservation of the angular momentum, one positive-helicity graviton is easy to be produced by two positive-helicity photons, and vice versa\footnote{One can actually check that this numerical discrepancy comes from the orthogonal relation of polarization vectors ~$e^{i \lambda}(\hat{\bm{k}})e{^{\lambda{'}*}}_i(\hat{\bm{k}}) = \delta_{\lambda\lambda'}$ .}.
 Therefore we can approximately get
\begin{align}
\left((2\pi)^3\delta^3(\bm{k} + \bm{k}')\right)^{-1}\langle\hat{h}_{S+}(\bm{k})\hat{h}_{S+}(\bm{k}')\rangle &\simeq \dfrac{4H^2\tau^2}{M_p^2}\int\dfrac{d\bm{p}}{(2\pi)^3}|e_{-}^i(\hat{\bm{k}})e^{+}_i(\hat{\bm{p}})|^2|e_{-}^j(\hat{\bm{k}})e^{+}_j(\hat{\bm{k}-\bm{p}})|^2 \notag \\
&\times\sum_{n=1}^\mathcal{N}\left|\int_{-\infty}^{0} d\tau_1\dfrac{G_k(\tau, \tau_1)}{a(\tau_1)}\hat{J}^{(n)}_+(\bm{p}, \bm{k}-\bm{p},\tau_1)\right|^2 \ , \\
\left((2\pi)^3\delta^3(\bm{k} + \bm{k}')\right)^{-1}\langle\hat{h}_{S-}(\bm{k})\hat{h}_{S-}(\bm{k}')\rangle &\simeq \dfrac{4H^2\tau^2}{M_p^2}\int\dfrac{d\bm{p}}{(2\pi)^3}|e_{+}^i(\hat{\bm{k}})e^{-}_i(\hat{\bm{p}})|^2|e_{+}^j(\hat{\bm{k}})e^{-}_j(\hat{\bm{k}-\bm{p}})|^2 \notag \\
&\times\sum_{m=1}^\mathcal{M}\left|\int_{-\infty}^{0} d\tau_1\dfrac{G_k(\tau, \tau_1)}{a(\tau_1)}\hat{J}^{(m)}_-(\bm{p}, \bm{k}-\bm{p},\tau_1)\right|^2 \ ,
\end{align}
where ~$G_k(\tau, \tau_1)$ ~is the retarded Green function associated with the homogeneous solution of ~$\hat{\psi}_\pm$ ~(denoted as \eqref{eq: Green}).
 Since different gauge fields are statistically uncorrelated with each other, ~$\langle\hat{h}_{S\pm}(\bm{k})\hat{h}_{S\pm}(\bm{k}')\rangle$ ~increase by the number of gauge fields ~$\mathcal{N}, \  \mathcal{M}$ .
 Note that in above calculations we do not include the commutation part of the source terms ~$\hat{J}^{(i)}_\pm$ .
 However, these simplifications are justified since mode functions of the produced gauge fluctuations are substantially real-valued, up to an irrelevant constant phase.
 Thus, defining the dimensionless power spectrum
\begin{align}
\langle\hat{h}_{\pm}(\bm{k})\hat{h}_{\pm}(\bm{k}')\rangle &\equiv (2\pi)^3\delta^3(\bm{k} + \bm{k}')\dfrac{2\pi^2}{k^3}\Delta^\pm_h(k) \ ,
\end{align}
we have the following tensor power spectra
%\begin{align}
%\Delta_h^\pm(k)  = \Delta_{hV}(k) + \Delta^\pm_{hS}(k) \simeq \dfrac{H^2}{\pi^2M_p^2}\left( 1 + \dfrac{2H^2}{M_p^2}f^+_h(\xi)e^{4\pi\xi} \right) \ ,
%\end{align}
\begin{align}
\Delta_h^+(k)  &= \dfrac{H^2}{\pi^2M_p^2}\left( 1 + \mathcal{N}\dfrac{2H^2}{M_p^2}f^+_h(\xi)e^{4\pi\xi} \right) \ , \\
\Delta_h^-(k) &= \dfrac{H^2}{\pi^2M_p^2}\left( 1 + \mathcal{M}\dfrac{2H^2}{M_p^2}f^+_h(\xi)e^{4\pi\xi} \right) \ , \\
f^+_h(\xi) &\simeq \dfrac{4.3 \times 10^{-7}}{\xi^6} \qquad (\xi \gtrsim 1) \ .
\end{align}
 Thus, both polarization modes of gravitons are amplified by the gauge fields and its chirality is characterized by the difference of ~$\mathcal{N}$ ~and ~$\mathcal{M}$ .
 So the amount of chirality ~$\chi$ ~is given by
\begin{equation}
\chi \equiv \dfrac{\Delta_h^+(k) - \Delta_h^-(k)}{\Delta_h^+(k) + \Delta_h^-(k)} \simeq \dfrac{\mathcal{N} - \mathcal{M}}{\mathcal{N} + \mathcal{M}} \qquad (\xi \gtrsim 1) \ .
\end{equation}
 Especially, in the case ~$\mathcal{N} = \mathcal{M}$ ~the provided tensor spectra are completely parity symmetric.
%Due to the increasing of ~$\xi$ ~during inflation, this spectra are naturally blue-tilted.
 On the contrary to the conventional predictions, these spectra could have the same amplitude and therefore predict no parity-violated phenomena through CMB and gravitational wave observations.

 At the end of this subsection, we also estimate the two-point correlation function of the curvature perturbation on uniform density hypersurfaces ~$\zeta = -H\delta\varphi/\dot{\bar{\varphi}}$.
 Not only tensor modes, but also the perturbation of inflaton ~$\delta\varphi$  ~is produced by the intrinsic inhomogeneities of Chern-Simons gauge interactions ~$\delta^{(i)}_{E B} \equiv \left( E^{(i)}_i B^{(i)}_i - \langle E^{(i)}_i B^{(i)}_i \rangle \right)|_{\delta\varphi=0}$.
 The equation of motion for ~$\delta\varphi$ ~reads\footnote{Note that we do not include the mass term of the fluctuation of inflaton and other cubic interactions via the scalar metric fluctuation because they are sub-dominant in the slow-roll expansion.}
\begin{align}
\ddot{\delta\varphi} + 3\beta H\dot{\delta\varphi} - a^{-2}\nabla^2\delta\varphi = \dfrac{1}{f}\left(\sum_{n=1}^\mathcal{N}\delta^{(n)}_{E B} - \sum_{m=1}^\mathcal{M}\delta^{(m)}_{E B} \right) \label{eq: deltavarphi} \ ,
\end{align}
where the additional friction term
\begin{align}
\beta \equiv 1 - \dfrac{1}{3fH}\left(\sum_{n=1}^\mathcal{N}\dfrac{\partial\langle E^{(n)}_i B^{(n)}_i \rangle}{\partial\dot{\bar{\varphi}}} - \sum_{m=1}^\mathcal{M}\dfrac{\partial\langle E^{(m)}_i B^{(m)}_i \rangle}{\partial\dot{\bar{\varphi}}}\right)
\end{align}
comes from the dependence of back-reaction ~$\langle E^{(i)}_i B^{(i)}_i \rangle$ ~on ~$\dot{\varphi}$ ~via ~$\xi$.
 Note that the back-reaction of the gauge field
\begin{equation}
\langle E^{(i)}_iB^{(i)}_i \rangle = -\dfrac{1}{a^4}\sum_{\lambda=\pm1}\int\dfrac{d^3k}{(2\pi)^3}\dfrac{k}{2} \lambda |A^{(i)}_\lambda|^2{'} \simeq -2.4\times10^{-4}\lambda\dfrac{H^4}{\xi^4}e^{2\pi\xi} \quad (\xi \gtrsim 1) \label{eq: back}
\end{equation}
depends on the sign of helicity of gauge field and therefore the contributions of each gauge field to the equation is the same sign.
 So ~$\beta$ ~is always positive.
 Here again, for ~$\delta\varphi$ ~we have two solutions ~$\delta\varphi = \delta\varphi_V + \delta\varphi_S$ ~from \eqref{eq: deltavarphi}.
 The homogeneous solution ~$\delta\varphi_V$ ~is an ordinal vacuum fluctuation so that ~$\langle \delta\varphi_{V\bm{k}}^2 \rangle^{1/2} = H/(2\pi)$ ~can be estimated.
 As to the peculiar solution ~$\delta\varphi_S$ ~sourced by the gauge field, we can quickly estimate \cite{Barnaby:2011qe}
\begin{equation}
\delta\varphi_S \sim \dfrac{\sum_{n=1}^\mathcal{N}\delta^{(n)}_{E B} - \sum_{m=1}^\mathcal{M}\delta^{(m)}_{E B}}{3f\beta H^2} \ .
\end{equation}
 As a result, defining the dimensionlsss power spectrum of the curvature perturbation
\begin{align}
\langle\hat{\zeta}(\bm{k})\hat{\zeta}(\bm{k}')\rangle \equiv (2\pi)^3\delta^3(\bm{k} + \bm{k}')\dfrac{2\pi^2}{k^3}\Delta_\zeta(k) \ ,
\end{align}
it is approximated as
\begin{equation}
\Delta_\zeta = \Delta_{\zeta v} + \Delta_{\zeta s} \simeq \left(\dfrac{H^2}{2\pi\dot{\bar{\varphi}}}\right)^2 + \left(\dfrac{\sigma}{3f\beta H\dot{\bar{\varphi}}}\right)^2 \label{eq: scalarpower} \ ,
\end{equation}
where ~$\sigma^2 \equiv \langle (\sum_{n=1}^\mathcal{N}\delta^{(n)}_{E B} - \sum_{m=1}^\mathcal{M}\delta^{(m)}_{E B})^2 \rangle$ ~is the variance of ~$E_iB_i$ .
 Since the different contribution to the two point functions of ~$\delta^{(i)}_{EB}, \ \delta^{(j)}_{EB} ~(i \neq j)$ ~adds incoherently ~$\langle \delta^{(i)}_{E B}\delta^{(j)}_{E B} \rangle = 0$ , we have
\begin{equation}
\sigma^2 = \sum_{i=1}^{\mathcal{N} + \mathcal{M}}\langle (\delta^{(i)}_{E B})^2 \rangle \simeq \sum_{i=1}^{\mathcal{N} + \mathcal{M}}\langle E^{(i)}_iB^{(i)}_i \rangle^2 \qquad (\xi \gtrsim 1) \ .
\end{equation}
 Therefore, contrary to the correlation function of tensor modes, that of scalar modes are suppressed by a factor ~$\mathcal{N}, \ \mathcal{M}$ ~at large ~$\xi$.

\subsection{Inflation with model-dependent and model-independent axion-gauge couplings}

 Next, we consider the case where Lagrangian density includes both a model-independent axion ~$\varphi_1$ ~and a model-dependent axion ~$\varphi_2$ ~coupled with two gauge groups.
 Here again, we consider the couplings of only Abelian gauge sectors
\begin{align}
\int d^4x\sqrt{-g_4}&\left[ \dfrac{M_p^2}{2}R
- \dfrac{1}{2}(\partial \varphi_1)^2 - \dfrac{1}{2}(\partial \varphi_2)^2 - V(\varphi_1, \varphi_2) - \sum_{n=1}^{\mathcal{N}}\dfrac{1}{4}F^{(n)}_{\mu\nu}F_{(n)}^{\mu\nu} - \sum_{m=1}^{\mathcal{M}}\dfrac{1}{4}F^{(m)}_{\mu\nu}F_{(m)}^{\mu\nu} \right. \notag \\
&\left. + \dfrac{1}{4}\dfrac{\varphi_1}{f_1}\left( \sum_{n=1}^{\mathcal{N}}F^{(n)}_{\mu\nu}\tilde{F}_{(n)}^{\mu\nu} + \sum_{m=1}^{\mathcal{M}}F^{(m)}_{\mu\nu}\tilde{F}_{(m)}^{\mu\nu} \right) +\dfrac{1}{4}\dfrac{\varphi_2}{f_2}\left( \sum_{n=1}^{\mathcal{N}}F^{(n)}_{\mu\nu}\tilde{F}_{(n)}^{\mu\nu} - \sum_{m=1}^{\mathcal{M}}F^{(m)}_{\mu\nu}\tilde{F}_{(m)}^{\mu\nu} \right) \right] \label{eq: model2} \ .
\end{align}
 %The point is that the combination of Chern-Simons terms coupled to both axions are different as we derived in the previous section.
 We assume that the trajectory of linear combination of two axions occurs inflation, analogous to the Kim-Nilles-Peloso (KNP) mechanism \cite{Kim:2004rp}.
 So, we diagonalize each variable as
\begin{align}
\varphi \equiv \dfrac{f_2}{\sqrt{f_1^2 + f_2^2}}\varphi_1 + \dfrac{f_1}{\sqrt{f_1^2 + f_2^2}}\varphi_2 \ , \qquad \tilde{\varphi} \equiv \dfrac{f_1}{\sqrt{f_1^2 + f_2^2}}\varphi_1 - \dfrac{f_2}{\sqrt{f_1^2 + f_2^2}}\varphi_2
\end{align}
and rewrite the axion-gauge couplings
\begin{align}
 \dfrac{1}{4}\dfrac{\varphi}{f}\left( \sum_{n=1}^{\mathcal{N}}F^{(n)}_{\mu\nu}\tilde{F}_{(n)}^{\mu\nu} + \dfrac{f_2^2 - f_1^2}{f_1^2 + f_2^2}\sum_{m=1}^{\mathcal{M}}F^{(m)}_{\mu\nu}\tilde{F}_{(m)}^{\mu\nu} \right) + \dfrac{1}{4}\dfrac{\tilde{\varphi}}{\tilde{f}}\sum_{m=1}^{\mathcal{M}}F^{(m)}_{\mu\nu}\tilde{F}_{(m)}^{\mu\nu} \label{eq: case2_2} \ ,
\end{align}
where we defined the following effective decay constants ~$f \equiv f_1f_2/\sqrt{f_1^2 + f_2^2} \ , \ \tilde{f} \equiv \sqrt{f_1^2 + f_2^2}/2$.
 Here, we assume that in the direction of ~$\varphi$ ~the inflation occurs and ~$\tilde{\varphi}$ ~has been stabilized on its potential during inflation.
 In addition to that, we impose the following hierarchy of decay constants ~$f_1 \gtrsim f_2$ .
 In this setup, the action \eqref{eq: model2} is effectively reduced to
\begin{align}
\int d^4x\sqrt{-g_4}&\left[ \dfrac{M_p^2}{2}R
- \dfrac{1}{2}(\partial \varphi)^2 - V(\varphi) - \sum_{n=1}^{\mathcal{N}}\dfrac{1}{4}F^{(n)}_{\mu\nu}F_{(n)}^{\mu\nu} - \sum_{m=1}^{\mathcal{M}}\dfrac{1}{4}F^{(m)}_{\mu\nu}F_{(m)}^{\mu\nu} \right. \notag \\
&\left. + \dfrac{1}{4}\dfrac{\varphi}{f}\left( \sum_{n=1}^{\mathcal{N}}F^{(n)}_{\mu\nu}\tilde{F}_{(n)}^{\mu\nu} - \alpha \sum_{m=1}^{\mathcal{M}}F^{(m)}_{\mu\nu}\tilde{F}_{(m)}^{\mu\nu} \right) \right] \label{eq: reducedmodel2} \ ,
\end{align}
%\begin{equation}
% \dfrac{1}{4}\dfrac{\varphi}{f}\left( F^{(2)}_{\mu\nu}\tilde{F}_{(2)}^{\mu\nu} - (1-2\alpha)F^{(1)}_{\mu\nu}\tilde{F}_{(1)}^{\mu\nu} \right) - V(\varphi) \label{eq: eff} \ ,
%\end{equation}
where ~$\alpha \equiv (f_1^2 - f_2^2)/(f_1^2 + f_2^2) > 0$ .
 Hence, as in the previous model, both polarization modes of the gauge fields are amplified due to the form of axion-gauge mixing, but the amplification of one helicity is characterized by ~$\alpha$.
 If the hierarchy is large ~$f_1^2 \gg f_2^2$, ~$\alpha$ ~can be approximated as ~$\alpha \simeq 1 - 2\epsilon$ ~with ~$\epsilon \equiv f_2^2/f_1^2$.

 Similar to the step in the previous subsection, we get the following tensor power spectra
\begin{align}
\Delta_h^+(k) &\simeq \dfrac{H^2}{\pi^2M_p^2}\left( 1 + \mathcal{N}\dfrac{2H^2}{M_p^2}f^+_h(\xi)e^{4\pi\xi} \right) \label{eq: plus} \ , \\
\Delta_h^-(k) &\simeq \dfrac{H^2}{\pi^2M_p^2}\left( 1 + \mathcal{M}\dfrac{2H^2}{M_p^2}f^+_h(\xi_\alpha)e^{4\pi\xi_\alpha} \right) \label{eq: minus} \ ,
\end{align}
and the scalar power spectrum
\begin{equation}
\Delta_\zeta \simeq \left(\dfrac{H^2}{2\pi\dot{\bar{\varphi}}}\right)^2 + \left(\dfrac{\sigma_\alpha}{3f\beta_\alpha H\dot{\bar{\varphi}}}\right)^2 \label{eq: scalarpower2} \ ,
\end{equation}
where ~$\xi_\alpha \equiv \alpha\xi$ ~and
\begin{align}
\beta_\alpha &\equiv 1 - \dfrac{1}{3fH}\left(\sum_{n=1}^{\mathcal{N}}\dfrac{\partial\langle E^{(n)}_i B^{(n)}_i \rangle}{\partial\dot{\bar{\varphi}}}(\xi) - \alpha\sum_{m=1}^{\mathcal{M}}\dfrac{\partial\langle E^{(m)}_i B^{(m)}_i \rangle}{\partial\dot{\bar{\varphi}}}(\xi_\alpha) \right), \\
\sigma_\alpha^2 &\equiv \sum_{n=1}^{\mathcal{N}}\langle (\delta^{(n)}_{E B}(\xi))^2 \rangle + \alpha^2\sum_{m=1}^{\mathcal{M}}\langle (\delta^{(m)}_{E B}(\xi_\alpha))^2 \rangle \ .
\end{align}
 In this case both polarization modes of graviton could be also enhanced and the chirality is parameterized by the number of gauge fields ~$\mathcal{N}, \ \mathcal{M}$ ~and the hierarchy of decay constants ~$\alpha$.
 At ~$\mathcal{N} = \mathcal{M}$ ~and large ~$\xi_{(\alpha)}$ , the amount of chirality of this spectra is approximately given by
\begin{equation}
\chi \simeq -\dfrac{\xi^{6} - \xi_\alpha^{6} e^{4\pi(\xi-\xi_\alpha)}}{\xi^{6} + \xi_\alpha^{6} e^{4\pi(\xi-\xi_\alpha)}} \ .
\end{equation}
 With a small ~$\epsilon$ , it is reduced to
\begin{equation}
\chi \sim (4\pi\xi - 6)\epsilon \ , \qquad \epsilon \lesssim \dfrac{1}{8\pi\xi} \label{eq: chira} \ .
\end{equation}
 It will be interesting to search for this small but non-negligible parity-violation in the tensor spectra with future experiments.
 In the next subsection, we show that the provided gravitational waves in this model are potentially testable with future gravitational wave experiments.

\subsection{Chiral blue tensor spectra}

 In this subsection, we numerically show the blue tensor spectra generated by the inflationary scenario presented in the last subsection.
 For the inflaton's potential, we adopt the KNP aligned cosine form
\begin{equation}
V(\varphi) = V_0\left[1 - \cos\left(\dfrac{\varphi}{h}\right)\right] \ , \qquad h \gtrsim M_p \ .
\end{equation}
 It is known that such a potential form could be realized by considering combinations of gaugino condensates and world-sheet instanton non-perturbative effects, or that of kinetic and Stuckelberg mixings of axion in string theory or string motivated supergravity \cite{Czerny:2014xja, Ben-Dayan:2014zsa, Long:2014dta, Gao:2014uha, Abe:2014pwa, Ali:2014mra, Shiu:2015uva, Ruehle:2015afa}.
 In order to agree with the constraint from the CMB observation, we keep the following condition \cite{Array:2015xqh}
\begin{equation}
\Delta_\zeta(k_{\text{CMB}}) \simeq 2 \times 10^{-9} \ , \qquad n_s \equiv \left.\dfrac{d\Delta_\zeta(k)}{d\ln k}\right|_{k=k_{\text{CMB}}} \simeq 0.96 \ , \qquad r \equiv \left.\dfrac{\Delta^+_h(k) + \Delta^-_h(k)}{\Delta_\zeta(k)}\right|_{k=k_{\text{CMB}}} \lesssim 0.07 \ ,
\end{equation}
where ~$k_{\text{CMB}} = 0.002\text{Mpc}^{-1}$ ~is the pivot scale which corresponds to ~$N(k_\text{CMB}) \equiv N_{\text{CMB}} \sim 50 - 60$ ~e-foldings before the end of inflation.
 Moreover, the gauge field contribution to the scalar modes is non-gaussian since it comes schematically from the product of two gaussian gauge fields, so that we have the following constraint on the small non-gaussianity in CMB scales \cite{Barnaby:2010vf}
\begin{equation}
\xi_{\text{CMB}} \equiv \xi(k=k_{\text{CMB}}) \lesssim 2.5 \ .
\end{equation}

\begin{figure}[h]
\begin{center}
\includegraphics[width=8cm,height=10cm,keepaspectratio]{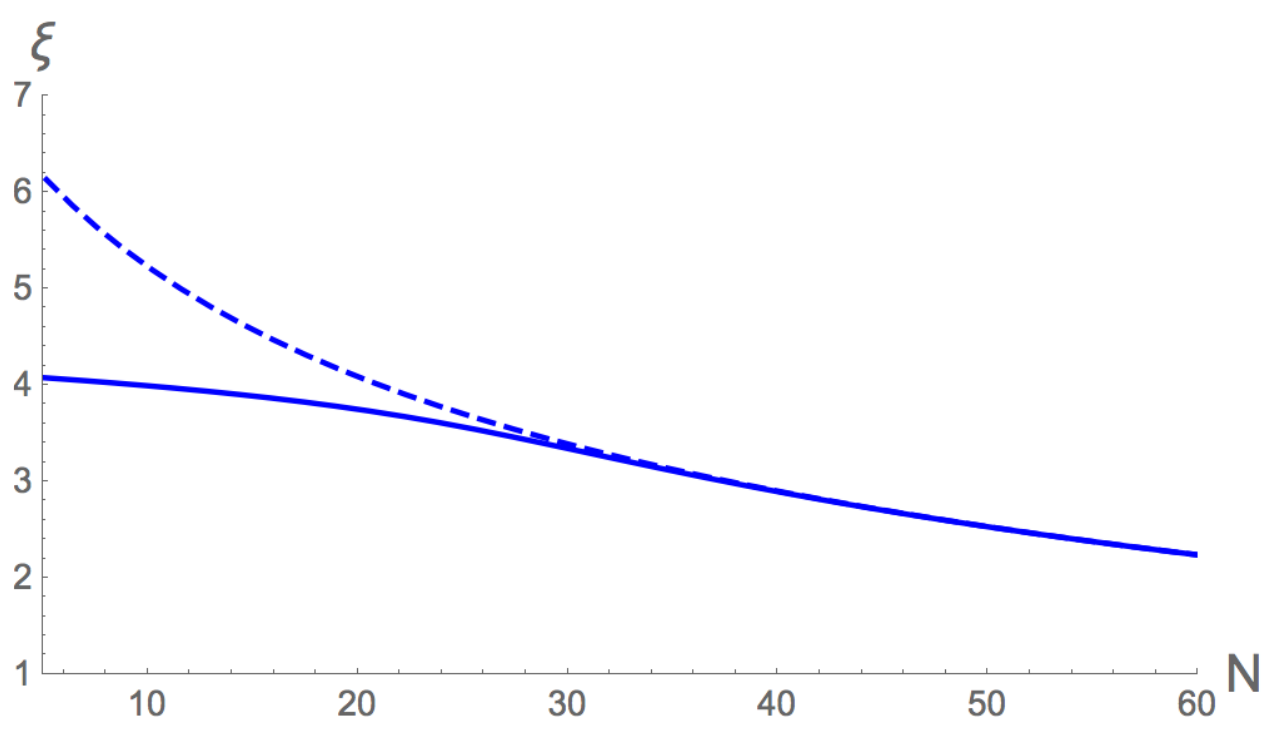}
\end{center}
\caption{We plotted the time evolution of ~$\xi$ ~as a function of the number of e-folds ~$N(t)$ ~with (solid blue line) and without (dashed blue line) including back-reaction of the gauge field on the motion of ~$\bar{\varphi}$ ~when ~$\xi_{\text{CMB}} \simeq 2.2$.
 Note that ~$N(t)$ ~counts from the end of inflation so that the time is moving from right to left in this figure.
 At the late time of inflation the effect of gauge field slows down the variation of ~$\bar{\varphi}$ ~and prolongs the inflationary phase.
 In this figure, we set ~$V_0 \simeq 1.56\times 10^{-9}M_p^4, \ f = 2\times10^{-2}M_p , \ h = 8M_p , \ \mathcal{N} = \mathcal{M} = 50, \ \epsilon = 10^{-2}$ .}
\label{fig: xi}
\end{figure}

 The background equation of motion for the axion reads
\begin{equation}
\ddot{\varphi} + 3H\dot{\varphi} + V_\varphi = \dfrac{1}{f}\left( \sum_{n=1}^{\mathcal{N}}\langle E^{(n)}_iB^{(n)}_i \rangle - \alpha \sum_{m=1}^{\mathcal{M}}\langle E^{(m)}_iB^{(m)}_i \rangle \right) \label{eq: backinflaton} \ ,
\end{equation}
where we introduce the back-reaction effect of gauge fields given by \eqref{eq: back}.
 We numerically plot the time evolution of ~$\xi$ ~in FIG.\ref{fig: xi}.
 We can see that in the late time of inflationary stage the axion moves more slowly due to the dissipation of gauge quanta.
 On the other hand, in the regime where the back-reaction effect of gauge fields is negligible, the evolution of ~$\xi$ ~is approximately given by
\begin{equation}
\xi \simeq \dfrac{M_p^2}{2fh}\sqrt{\dfrac{\cos^2(\varphi_{\text{end 0}}/2h)\exp(-M_p^2N_0/h^2)}{1-\cos^2(\varphi_{\text{end 0}}/2h)\exp(-M_p^2N_0/h^2)}} \ ,
\end{equation}
where ~$\varphi_{\text{end} 0}$ ~and ~$N_0$ ~denote the value of inflaton at the end of inflation and the number of e-folds, both without including back-reaction of gauge fields.
 Then we write the number of e-foldings as a function of frequency
\begin{equation}
N(k=2\pi f) \equiv \ln\left(\dfrac{a_{\text{end}}}{a_k}\right) = N_{\text{CMB}} + \ln\left(\dfrac{k_{\text{CMB}}}{0.002\text{Mpc}^{-1}}\right) - \ln\left(\dfrac{f}{3\times 10^{-18}\text{Hz}}\right)
\end{equation}
\begin{figure}[h]
\begin{center}
\includegraphics[width=15cm,height=15cm,keepaspectratio]{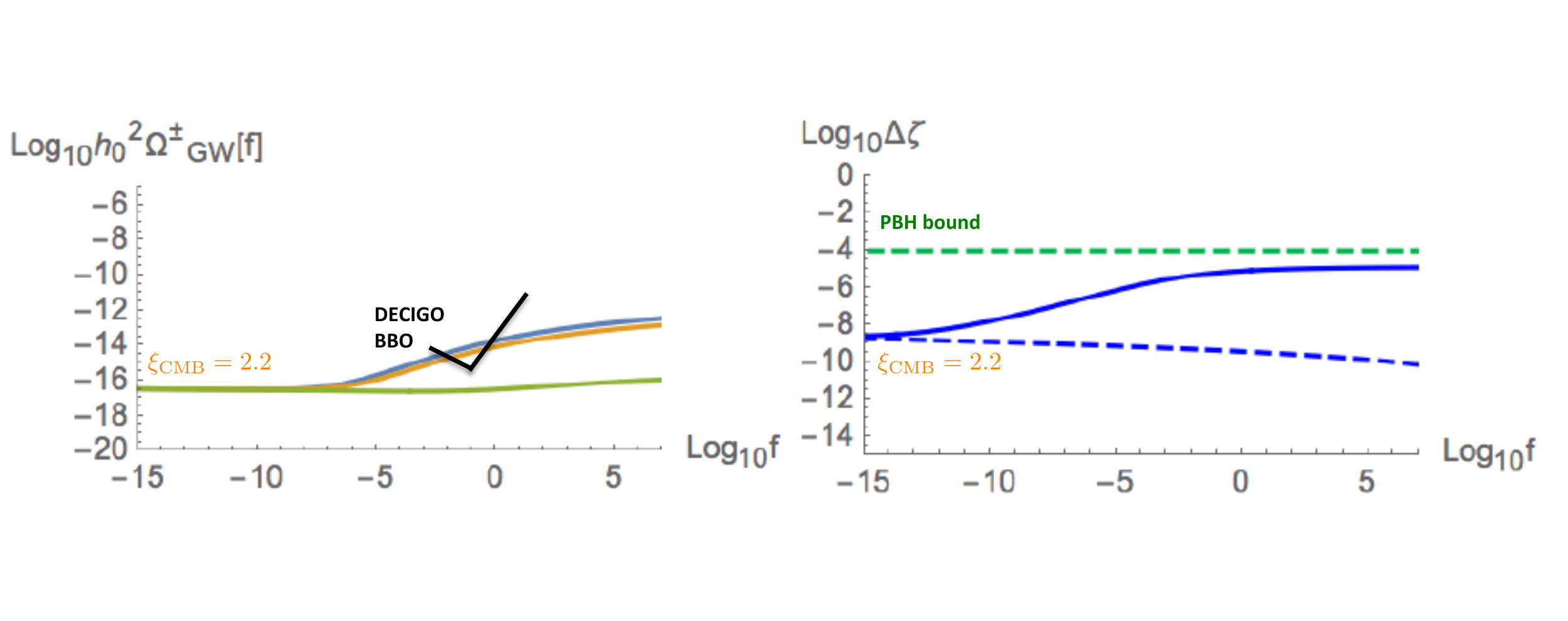}
\end{center}
\caption{Plot of the tensor and the scalar power spectrum when ~$\xi_{\text{CMB}} \simeq 2.2$.
 In the left figure, we depict the present gravitational energy intensity of plus mode (blue line), minus mode with ~$\epsilon = 10^{-2}$ (orange line) and ~$\epsilon = 10^{-1}$ (green line), which are naturally blue-shaped due to the increase of velocity of axion and potentially testable by future space-based gravitational interferometer DECIGO or BBO.
 In the right figure, we plot the power spectrum of curvature perturbation with ~$\epsilon = 10^{-2}$ ~(solid blue line).
 The dashed blue line represents the power spectrum of vacuum fluctuations.
 We can see that it satisfies an upper bound on the non-detection of PBHs (green dashed line). 
 In both figures, we set ~$V_0 \simeq 1.56\times 10^{-9}M_p^4, \ f = 2\times10^{-2}M_p , \ h = 8M_p, \ \mathcal{N} = \mathcal{M} = 50$ ~in this plot.}
\label{fig: mainresult}
\end{figure}
and depict the evolution of the tensor spectra \eqref{eq: plus} \eqref{eq: minus} in the left-side of FIG.\ref{fig: mainresult}.
 Here, we rewrite the tensor spectra as the present energy intensity of gravitational waves ~$\Omega^\pm_{\text{GW}}(f)$ ~expressed by ~$\Omega^\pm_{\text{GW}} \simeq 10^{-6}h_0^{-2}\Delta^\pm_h$ ~with the dimensionless Hubble parameter ~$h_0 \simeq 0.7$ \cite{Smith:2005mm}.
 We can see that these amplitudes are growing on small scales due to the increase of ~$\xi$ and testable with DECIGO or BBO gravitational wave experiments when ~$\mathcal{N} = \mathcal{M} = 50, \ \epsilon \sim 10^{-2}$.
 In this case, from \eqref{eq: chira} the amount of chirality is about ~$\chi \sim 0.5$ .
 If sizable blue tensor spectra are produced, we can test the parity-violation down to ~$\chi = 0.08$ ~by these experiments \cite{Seto:2006dz}.
 Therefore, we have a chance to search for this small chirality of primordial tensor spectra in detail.

 Simultaneously, however, we should also take care of the amplification of curvature perturbation, leading to the overproduction of PBHs in small scales.
 It is known that when the curvature perturbation exceeds a critical value ~$\zeta \gtrsim \zeta_c \sim 1$, at horizon re-entry after inflation it leads to the black hole formation.
 We can represent this probability in terms of the variance ~$\langle \zeta_c^2 \rangle$ ~and constrain the amplitude of the power-spectrum of curvature perturbation.
 In our case, the curvature perturbation sourced by gauge fields is non-gaussian and is schematically written by ~$\zeta = g^2 - \langle g^2 \rangle$, where ~$g$ ~is the Gaussian distribution field.
 So its variance is roughly estimated as ~$\Delta_{\zeta} \sim \langle g^2 \rangle^2$.
 The bound on the distribution of non-gaussian ~$\zeta$ ~has already been studied and estimated by \cite{Linde:2012bt, Lyth:2012yp}
\begin{equation}
\Delta_{\zeta_c} = \mathcal{O}(10^{-4}) - \mathcal{O}(10^{-3}) \ .
\end{equation}
 In the right-side of FIG.\ref{fig: mainresult}, we plot the time evolution of the power-spectrum of curvature perturbation given by \eqref{eq: scalarpower2} and a rough upper bound on the constraint of PBHs.
 We can see that in this figure our model avoids the overproduction of curvature perturbation and satisfies the upper bound on the non-detection of PBHs during inflation.

\section{Summary \& Discussion}

 In this paper, we studied the new feature of generating primordial blue tensor spectra from the inflationary axion-gauge interactions, whose coupling forms are inspired by the dimensional reduction of string theory.
%explored the axionic inflationary dynamics coupled with two gauge groups, inspired by the Green-Schwarz mechanism in string theory, and studied the time evolution of perturbations sourced by the gauge field during inflation.
 In the string theory, axions arise from the compactifications of the form fields, which couple with the gauge field in order to cancel the gauge anomaly in ten dimension.
 The important property is that there are two types of axion, dubbed the model-independent axion and the model-dependent axion, which acquire the different sign combination of the coupled gauge groups through the generalized Green-Schwarz mechanism.
%one is lead to the component of the form field tangent to our four-dimensional space-time and is called the model-independent axion, the other is its component tangent to the compact manifold and is called the model-dependent axion. 
% As an universal property, the model-dependent axion acquires the coupling of Yang-Mills Chern-Simons terms through the generalized Green-Schwarz counter terms.
% When axion couples with two gauge groups, with opposite sign each other. 
 %More precisely, we considered the inflationary scenarios occurred by two types of axion, originating from the compactifications of the form fields in string theory, which couple with two gauge groups through the anomaly cancellation mechanism in ten-dimensional theory.
 Considering this fact, we constructed the inflationary models where the model-independent or the model-dependent axion plays a role of inflaton, both of which couple with Abelian gauge fields with same or opposite signs each other. 

 The main prediction of this paper is that several types of blue tenor spectra are provided due to the amplifications of each gauge group.
 If the couplings of the model-dependent axion to more than two gauge fields with different signs are included, both polarization modes of gauge fields experiences a tachyonic instability and enhances both helicity of metric tensor modes during inflation.
 As a result, in the case where only the model-dependent axion plays a role of an inflaton, the totally parity-conserved tensor power spectra could be provided.
 This prediction is completely different from the previous results since a sizable amplitude of tensor power spectrum is totally parity-violated in small scales.
 On the other hand, when both types of axion play a role of an inflaton, we developed the possibility of generating the blue tensor spectra whose chirality is parameterized by the hierarchy of their decay constants.
 As an example, we described the dynamics of generating tensor power spectrum in the case of a cosine-type potential and found that the provided blue tensor spectra with slightly parity-violated have the possibility to detect the future gravitational wave experiments such as DECIGO or BBO project.
 %If sizable blue tensor spectra are produced, we can test the parity-violation up to ~$\chi = 0.08$ ~by these experiments \cite{Seto:2006dz}.
 %Therefore, we have a chance to search for this small chirality of primordial tensor spectra in detail.
 As to the dynamics of scalar curvature perturbation, we checked the time evolution of its two correlation function and showed that it safely avoids the overproduction of PBHs in small scales.
 In this work, however, we have not estimated the motion of the inflaton with back-reaction of gauge fields self-consistently so that the careful treatment of the evolution of curvature perturbation will be actually needed near the end of inflation where back-reaction is strongly dominated \cite{Cheng:2015oqa, Notari:2016npn}.

 It would be instructive to consider how unique our prediction is.
 It seems that similar results could appear when two axions couple with the gauge fields with the same signs and roll down on their potentials in an opposite direction each other.
 In this case, however, we will need fine-tuning for each parameter of axion's velocity or the number of gauge fields to realize our result because in the equation of motion for each gauge field the ~$\xi$ ~parameter is not in common.
 In addition to that, there is another mechanism of generating chiral primordial gravitational waves \cite{Satoh:2007gn, Satoh:2008ck, Takahashi:2009wc, Wang:2012fi, Cai:2016ihp}, so it is worth to consider whether we can discriminate between these theories through the observation of parity-violated tensor spectrum.
 Moreover, it is intriguing to study the case where axions play a role of spectator fields during inflation.
 We expect that in this case the similar result of this work would be also realized.
 Finally, in this work we search for the dynamics of Abelian gauge fields for simplicity and neglect the effect of higher interaction terms of non-Abelian gauge fields.
 Hence, it will be important to explore the dynamics of axions coupled to non-Abelian gauge sectors during inflation.
 We leave these issues in our future work.

\section*{Acknowledgements}

 We thank to Jiro Soda, Toshifumi Noumi and Hajime Otsuka for fruitful discussion and advices.
 This work was supported by JSPS KAKENHI Grant Number 15J01345. 

\appendix

\section{Short review of the particle production of U(1) gauge field}

 In this appendix, we review the mechanism of the particle production of U(1) gauge field ~$A_\mu$ ~coupled to the axion ~$\varphi$ ~which plays a role of inflaton.
 We consider the following action
\begin{equation}
S = \int d^4x \sqrt{-g}\left[ \dfrac{M_p^2}{2}R -\dfrac{1}{2}(\partial\varphi)^2 - V(\varphi) - \dfrac{1}{4}F_{\mu\nu}F^{\mu\nu} + \dfrac{1}{4}\dfrac{\varphi}{f}F_{\mu\nu}\tilde{F}^{\mu\nu} \right] \ ,
\end{equation}
where ~$R$ ~is the Ricci scalar (~$g = \det g_{\mu\nu}$ ~is a determinant of a metric), ~$F_{\mu\nu} = \partial_\mu A_\nu - \partial_\nu A_\mu$ ~is the field strength of U(1) gauge field.
 As to the gauge fixing, we chose temporal gauge ~$A_0 = 0$ ~and Coulomb gauge ~$\partial_iA_i = 0$ .
 Its Hodge duals are defined by ~$\tilde{F}^{\mu\nu} = \sqrt{-g}\epsilon^{\mu\nu\rho\sigma}F_{\rho\sigma}/2!$ , where ~$\epsilon^{\mu\nu\rho\sigma}$ ~is the antisymmetric Levi-Civita tensor which satisfies ~$\epsilon^{0123} = g^{0\mu}g^{1\nu}g^{2\rho}g^{3\sigma}\epsilon_{\mu\nu\rho\sigma} = 1/g$ .
 As to the metric, we assume a spatially flat FLRW metric
~$ds^2 = -dt^2 + a(t)^2d\bm{x}^2 = a(\tau)^2[-d\tau^2 + d\bm{x}^2]$
~where ~$a$ ~is the scale factor as a function of the cosmic time ~$t$ ~or the conformal time ~$\tau$ .

 We firstly solve the equation of motion for the gauge field in a homogeneous background, assuming that back-reaction on the inhomogeneous perturbation is negligible.
 It is given by
\begin{equation}
A_i{''} - \nabla^2A_i - \dfrac{\bar{\varphi}'}{f}\epsilon_{ijk}\partial_jA_k = 0 \label{eq: A0} \ ,
\end{equation}
where the prime means the derivative with respect to ~$\tau$ ~and  ~$\bar{\varphi} = \bar{\varphi}(t)$ ~is a background inflaton field.
 In order to describe the production of quanta of gauge fields, we promote the classical field ~$A_i$ ~to an operator expansion
\begin{align}
A_i(\tau, \bm{x}) &= \sum_{\lambda = \pm}\int\dfrac{d\bm{k}}{(2\pi)^3}\hat{A}_{\lambda}(\bm{k}, \tau)e{^\lambda}_i(\hat{\bm{k}})e^{i\bm{k}\cdot\bm{x}} \label{eq: Fouriergauge} \ , \\
\hat{A}_{\lambda}(\bm{k}, \tau)&= a_{\lambda\bm{k}}A_{\lambda}(k, \tau) + a{^{\dagger}}_{\lambda-\bm{k}}A^{*}_{\lambda}(k, \tau) \ ,
\end{align}
where ~$a_{\lambda\bm{k}} , ~a{^{\dagger}}_{\lambda-\bm{k}}$ ~are annihilation and creation quantum operators which obey the following commutation relation
~$
[a_{\lambda\bm{k}} , a{^{\dagger}}_{\lambda'-\bm{k}'}] = \delta_{\lambda\lambda'}(2\pi)^3\delta^3(\bm{k} + \bm{k}') \ .
$~
 The polarization vector ~$e{^\pm}_i(\hat{\bm{k}}) = e{^{\mp*}}_i(\hat{\bm{k}})$ ~satisfies the normalization-orthogonal condition ~$e^{i \lambda}(\hat{\bm{k}})e{^{\lambda{'}*}}_i(\hat{\bm{k}}) = \delta_{\lambda\lambda'}$ , ~the transverse property ~$k^i e{^\pm}_i(\hat{\bm{k}}) = 0$ ~and the circular polarization state defined by ~$i\epsilon_{ijl}~k^je^{l \pm}(\hat{\bm{k}}) = \pm k e{^{\pm}}_i(\hat{\bm{k}})$ .
 Substituting these relations into \eqref{eq: A0}, we have the equation of motion for each mode function
\begin{equation}
\left[ \partial_{\tau}^2 + k^2 \pm\dfrac{2k\xi}{\tau}\right]A_{\pm} = 0 \ , \qquad \xi \equiv \dfrac{\dot{\bar{\varphi}}}{2fH} \quad \left( ``\cdot" = \dfrac{d}{dt} \right) \label{eq: A1} \ .
\end{equation}
 Here, we defined a new parameter ~$\xi$ ~which characterizes the production rate of the gauge field as we will show later.
 Since the time variation of ~$\xi$ ~is slow-roll suppressed, we can approximate it as a constant.
 For our convenience, we choose ~$\xi$ ~as a positive value.

 Let us analyze \eqref{eq: A1}.
 The exact solution is given by the Whittaker function
~$
A_{\pm} = e^{\pm\pi\xi/2}W_{\mp i\xi, 1/2}(2i k\tau)/\sqrt{2k} \ ,
$~
where we assume that the gauge field is initially in the adiabatic vacuum ~$A_{\pm} \rightarrow e^{-i k\tau}/\sqrt{2k} ~(-k\tau \rightarrow \infty)$.
 While the minus mode ~$A_-$ ~is not amplified, we can see that the plus mode $A_+$ experiences a tachyonic instability for ~$-k\tau < 2\xi$.
 We are interested in the parameter region ~$\xi \gtrsim 1$ , so that around the horizon-crossing the gauge field can be substantially approximated as the following real value
\begin{align}
A_+ &\simeq \dfrac{1}{\sqrt{2k}}\left(\dfrac{-k\tau}{2\xi}\right)^{1/4}e^{\pi\xi-2\sqrt{-2\xi k\tau}} \equiv \dfrac{1}{\sqrt{2k}}\mathcal{A}(\xi, -k\tau) \label{eq: A} \ , \\
A_+{'} &\simeq \sqrt{\dfrac{\xi}{-\tau}}\mathcal{A}(\xi, -k\tau) \label{eq: A'}
\end{align}
in the following time interval ~$(8\xi)^{-1} \lesssim -k\tau \lesssim 2\xi$ .
 It is enhanced by a factor ~$e^{\pi\xi}$ ~and this amplification affects the production of tensor and scalar perturbations.
 In the super-horizon regime ~$-k\tau \rightarrow 0$ , the enhanced mode ~$A_+$ ~settles in a constant value ~$e^{\pi\xi}/(2\sqrt{\pi k \xi})$, so the energy density of gauge fields dilutes due to the expansion of the universe.
 Thus, in the region ~$(8\xi)^{-1} \lesssim -k\tau \lesssim 2\xi$ ~the gauge field makes the most contribution to the physical enhancement of fluctuations.

 The enhancement of gauge field could also affect the background dynamics.
 For the Friedmann equation and the equation of motion for the axion ~$\varphi= \varphi(t)$ , they read
\begin{equation}
3M_p^2H^2 = \dfrac{1}{2}\dot{\varphi}^2 + V(\varphi) + \dfrac{1}{2}\langle E_i^2 + B_i^2 \rangle \ ,
\end{equation}
\begin{equation}
\ddot{\varphi} + 3H\dot{\varphi} + V_\varphi = \dfrac{1}{f}\langle E_iB_i \rangle \label{eq: inflaton} \ ,
\end{equation}
%\begin{equation}
%\varphi'' + 2\mathcal{H}\varphi' - \nabla^2\varphi + a^2V_\varphi = \dfrac{a^2}{f}\left[ (E^a_iB^a_i)_{(F)} - (1-2\alpha)(E^a_iB^a_i)_{(G)} \right] \ ,
%\end{equation}
where we defined the electric and magnetic components of the gauge field as
\begin{equation}
E_i \equiv -\dfrac{1}{a^2}A_i' \ , \qquad B_i \equiv \dfrac{1}{a^2}\epsilon_{ijk}\partial_j A_k \label{eq: elemag} \ .
\end{equation}
 In above equations, we introduce the back-reaction effect of the amplification of gauge field.
%expected by the adiabatic vacuum state ~$a_{\lambda\bm{k}}\ket0 = 0$ .
 They are given by
\begin{align}
\left\langle E_iB_i \right\rangle &= -\dfrac{1}{a^4}\sum_{\lambda=\pm1}\int\dfrac{d^3k}{(2\pi)^3}\dfrac{k}{2} \lambda |A_\lambda|^2{'} \simeq -H^4\int_0^{2\xi}\dfrac{x^3dx}{4\pi^2}\sqrt{\dfrac{2\xi}{x}}|\mathcal{A}|^2 \label{eq: back1} \ , \\
\dfrac{1}{2}\langle E_i^2 + B_i^2 \rangle &= \dfrac{1}{a^4}\sum_{\lambda=\pm1}\int\dfrac{d^3k}{2(2\pi)^3}\left( |A_\lambda{'}|^2 + k^2|A_\lambda|^2\right) \simeq H^4\int_0^{2\xi}\dfrac{x^3dx}{4\pi^2}\left( \dfrac{\xi}{x} + \dfrac{1}{2} \right)|\mathcal{A}|^2 \label{eq: back2} \ ,
\end{align}
where ~$x \equiv -k\tau$ ~and we assume that the UV-divergence of the vacuum mode at ~$x \gg 2\xi$ ~is renormalized away.
 Note that we can extend the integration of ~$\mathcal{A}$  ~to the super-horizon limit since it is actually a good approximation of the exact solution.
 At large ~$\xi$, these expectation values are approximately given by \cite{Anber:2012du}
\begin{equation}
\left\langle E_iB_i \right\rangle \simeq -2.4\times10^{-4}\dfrac{H^4}{\xi^4}e^{2\pi\xi} \ , \quad \dfrac{1}{2}\langle E_i^2 + B_i^2 \rangle \simeq 1.4\times10^{-4}\dfrac{H^4}{\xi^3}e^{2\pi\xi} \ .
\end{equation}
 Therefore, we need the following constraint to neglect back-reaction on the inflationary dynamics
\begin{equation}
\dfrac{\left\langle E_iB_i \right\rangle}{3fH\dot{\bar{\varphi}}} \ll 1 \ , \qquad \dfrac{\langle E_i^2 + B_i^2 \rangle}{6H^2} \ll 1 \label{eq: scalar} \ .
\end{equation}
 Of these conditions the first one is always stronger than the second one in our parameter region.

\subsection{Tensor mode amplification}

 Here, we present the mechanism of providing the metric tensor modes sourced by the gauge field.
 We use the inverse decay method in this calculation\footnote{It is known that this method derives almost the same result when we use the in-in method \cite{Barnaby:2011qe}.}.
 The tensor modes ~$h_{ij}$ ~are included in the spatial components of the metric as ~$g_{ij} = a^2(\delta_{ij} + h_{ij})$ ~with ~$\partial_i h_{ij} = h_{ii} = 0$ .
 Defining a new canonical variable ~$\psi_{ij} = aM_ph_{ij}/2$, ~the equation of motion for ~$\psi_{ij}$ ~is obtained from the Einstein equation
%\begin{equation}
%h_{ij}'' + 2\mathcal{H}h_{ij}' - \nabla^2h_{ij} = -\dfrac{2a^2}{M_p^2}\Pi_{ij}^{lm}\{(E^a_lE^a_m + B^a_lB^a_m)_{(F)} + (E^a_lE^a_m + B^a_lB^a_m)_{(G)} \} \ ,
%\end{equation}
\begin{equation}
\left[\partial_\tau^2 - \nabla^2 - \dfrac{2}{\tau^2} \right]\psi_{ij} = -\dfrac{a^3}{M_p}\Pi_{ij}^{lm}(E_lE_m + B_lB_m) \ ,
\end{equation}
where ~$\Pi_{ij}^{lm}$ ~is the transverse-traceless projector given by
\begin{equation}
\Pi_{ij}^{lm} \equiv \Pi_i^l\Pi_j^m - \dfrac{1}{2}\Pi_{ij}\Pi^{lm} \ , \qquad \Pi_{ij} \equiv \delta_{ij} - \dfrac{\partial_i\partial_j}{\nabla^2} \ .
\end{equation}
 In order to get the equation in Fourier modes, we go to the momentum space of ~$\psi_{ij}$ :
\begin{align}
\psi_{ij} &= \int\dfrac{d\bm{k}}{(2\pi)^3}\hat{\psi}_{ij}(\bm{k})e^{i\bm{k}\cdot\bm{x}} \notag \\
&= \sum_{\lambda = \pm}\int\dfrac{d\bm{k}}{(2\pi)^3}e^\lambda_{ij}(\hat{\bm{k}})\hat{\psi}_\lambda(\bm{k})e^{i\bm{k}\cdot\bm{x}} \label{eq: Fouriertensor} \ ,
\end{align}
where we defined the polarization tensor ~$e^\pm_{ij}(\hat{\bm{k}}) \equiv e^\pm_i(\hat{\bm{k}})e^\pm_j(\hat{\bm{k}})$ ~so that the following relationships ~$\hat{\psi}_\pm(\bm{k}) = e^{ij}_{\mp}(\hat{\bm{k}})\hat{\psi}_{ij}(\bm{k}), \ e^{ij}_\pm \Pi_{ij}^{lm} = e^{lm}_\pm$ ~are held.
 Substituting these relations, we get
\begin{align}
\left[\partial_\tau^2 + k^2 - \dfrac{2}{\tau^2} \right]\hat{\psi}_\pm(\bm{k}) = -\dfrac{a^3}{M_p}e^{lm}_\mp(\hat{\bm{k}})\int\dfrac{d\bm{p}}{(2\pi)^3}& \left(\hat{E}_l(\bm{p})\hat{E}_m(\bm{k} - \bm{p}) + \hat{B}_l(\bm{p})\hat{B}_m(\bm{k} - \bm{p})\right) \label{eq: tensor} \ ,
\end{align}
where
\begin{equation}
\hat{E}_i(\bm{p}) = -\sum_{\lambda=\pm}\dfrac{1}{a^2}\hat{A}_\lambda{'} e{^\lambda}_i(\hat{\bm{p}}) \ , \qquad \hat{B}_i(\bm{p}) = \sum_{\lambda=\pm}\dfrac{\lambda k}{a^2}\hat{A}_\lambda e{^\lambda}_i(\hat{\bm{p}}) \ .
\end{equation}
 In \eqref{eq: tensor}, we can separate the solution of ~$\hat{\psi}_\pm$ ~into two parts
~$
\hat{\psi}_\pm = \hat{\psi}_{V} + \hat{\psi}_{S\pm}
$~:
 the first term ~$\hat{\psi}_{V}$ ~is a homogeneous solution which comes from vacuum fluctuations of quasi de Sitter space-time, so its mode function obeys the following Bunch-Davies solution
\begin{equation}
\psi_{V} = \dfrac{1}{\sqrt{2k}}\left( 1 - \dfrac{i}{k\tau} \right)e^{-ik\tau} \ .
\end{equation}
 On the other hand, the second term ~$\hat{\psi}_{S\pm}$ ~is a peculiar solution sourced by the gauge field in the r.h.s. of \eqref{eq: tensor}, so that it is given by
\begin{align}
\hat{\psi}_{S\pm}(k, \tau) &= -\dfrac{1}{M_p}\int_{-\infty}^{0} d\tau_1\dfrac{G_k(\tau, \tau_1)}{a(\tau_1)}\int\dfrac{d\bm{p}}{(2\pi)^3}e^{lm}_\mp(\hat{\bm{k}}) \notag \\
&\times \left[ \left(\hat{A}_+{'}(\bm{p})\hat{A}_+{'}(\bm{k}-\bm{p}) + p|\bm{k} - \bm{p}|\hat{A}_+(\bm{p})\hat{A}_+(\bm{k}-\bm{p})\right) e^+_l(\hat{\bm{p}})e^+_m(\hat{\bm{k}-\bm{p}}) \right. \notag \\
&\left. + \left(\hat{A}_-{'}(\bm{p})\hat{A}_-{'}(\bm{k}-\bm{p}) + p|\bm{k} - \bm{p}|\hat{A}_-(\bm{p})\hat{A}_-(\bm{k}-\bm{p})\right) e^-_l(\hat{\bm{p}})e^-_m(\hat{\bm{k}-\bm{p}}) \right] \label{eq: pe} \ ,
\end{align}
where
\begin{align}
G_k(\tau, \tau_i) &= i~\Theta(\tau - \tau_i)\left(\psi_V(k, \tau)\psi^*_V(k, \tau_i) - \psi^*_V(k, \tau)\psi_V(k, \tau_i)\right) \notag \\
&= i~\Theta(\tau - \tau_i)\left[ \dfrac{\tau - \tau_i}{k^2\tau\tau_i}\cos[k(\tau-\tau_i)] - \dfrac{k^2\tau\tau_1 + 1}{k^3\tau\tau_1}\sin[k(\tau-\tau_i)] \right] \label{eq: Green}
\end{align}
is the retarded Green function associated with the homogeneous solution of \eqref{eq: tensor}.

 Let us analyze the two-point correlation function of ~$\hat{\psi}_{S\pm}$ .
 Firstly, in \eqref{eq: pe} we can approximate ~$\hat{A}_- = 0$ ~in the integration since the most contribution part of the integration comes from the enhancement of ~$\hat{A}_+$ ~around horizon-crossing, so we ignore the terms in the third line of \eqref{eq: pe}.
 Then after renormalization of UV part, we can reduce the integration of time interval to the region where the particle production of the gauge field takes place.
 Therefore, rewriting ~$\hat{\psi}_{S\pm}$ ~in ~$\hat{h}_{S\pm} = 2\hat{\psi}_{S\pm}/(aM_p)$ ~we can find
%\begin{align}
%\langle\hat{\psi}_{I\pm}(\bm{k})\hat{\psi}_{I\pm}(\bm{k}')\rangle &\simeq (2\pi)^3\delta^3(\bm{k} + \bm{k}')\dfrac{k^3}{M_p^2}\int d\tau_1\dfrac{G_k(\tau, \tau_1)}{a(\tau_1)}\int d\tau_2\dfrac{G_k(\tau, \tau_2)}{a(\tau_2)}\int\dfrac{d\bm{p_*}}{(2\pi)^3} \notag \\
%&\times\left[ |e_{\mp}^l(\hat{\bm{k}})e^+_l(\hat{\bm{p}})|^2|e_{\mp}^m(\hat{\bm{k}})e^+_m(\hat{\bm{k}-\bm{p}})|^2e^{4\pi\xi}e^{-2\sqrt{2\xi}\left(\sqrt{p_*} + \sqrt{|1-p_*|}\right)\left(\sqrt{-k\tau_1} + \sqrt{-k\tau_2}\right)}\dfrac{k^3\sqrt{p_*|1-p_*|}\sqrt{\tau_1\tau_2}}{4\xi} \right. \notag \\
%&\left.\times\left( \dfrac{4\xi^2}{k^2\tau_1\tau_2} + \dfrac{4\xi\sqrt{p_*|1-p_*|}}{-k\tau_1} + p_*|1-p_*| \right) + |e_{\mp}^l(\hat{\bm{k}})e^-_l(\hat{\bm{p}})|^2|e_{\mp}^m(\hat{\bm{k}})e^-_m(\hat{\bm{k}-\bm{p}})|^2 \right. \notag \\
%&\left. \times e^{4\pi\xi_\alpha}e^{-2\sqrt{2\xi_\alpha}\left(\sqrt{p_*} + \sqrt{|1-p_*|}\right)\left(\sqrt{-k\tau_1} + \sqrt{-k\tau_2}\right)}\dfrac{k^3\sqrt{p_*|1-p_*|}\sqrt{\tau_1\tau_2}}{4\xi_\alpha} \right. \notag \\
%&\left.\times\left( \dfrac{4\xi_\alpha^2}{k^2\tau_1\tau_2} + \dfrac{4\xi_\alpha\sqrt{p_*|1-p_*|}}{-k\tau_1} + p_*|1-p_*| \right) \right] \ ,
%\end{align}
\begin{equation}
\langle\hat{h}_{S\pm}(\bm{k})\hat{h}_{S\pm}(\bm{k}')\rangle \simeq (2\pi)^3\delta^3(\bm{k} + \bm{k}')\dfrac{4H^4}{k^3M_p^4} f^\pm_h(\xi)e^{4\pi\xi} \ ,
\end{equation}
where we denoted a numerical factor
\begin{align}
f^\pm_h(\xi) &\equiv \int\dfrac{d\bm{p_*}}{(2\pi)^3}|e_{\mp}^l(\hat{\bm{k}})e^{+}_l(\hat{\bm{p}})|^2|e_{\mp}^m(\hat{\bm{k}})e^{+}_m(\hat{\bm{k}-\bm{p}})|^2 \dfrac{\sqrt{p_*|\hat{\bm{k}} - \bm{p}_*|}}{4\xi} \notag \\
&\times \left[ \int^0_\infty dx_1 ~x_1^{3/2} \left(\cos x_1 - \dfrac{\sin x_1}{x_1} \right)\left( \dfrac{2\xi}{x_1} + \sqrt{p_*|\hat{\bm{k}} - \bm{p}_*|} \right)e^{-2\sqrt{2\xi}\left(\sqrt{p_*} + \sqrt{|\hat{\bm{k}} - \bm{p}_*|}\right)\sqrt{x_1}} \right]^2 \label{eq: fxi} \ .
\end{align}
 Here $x_1 = -k\tau, \ \bm{p}_* \equiv \bm{p}/|\bm{k}|$ ~and we used the following relationship ~$|e_{\mp}^l(\hat{\bm{k}})e^\lambda_l(\hat{\bm{p}})|^2 = \left( 1 \pm\lambda \hat{\bm{k}}\cdot\hat{\bm{p}} \right)^2/4$ ~in above expressions.
 Although we used the approximate solution \eqref{eq: A} to the whole integrand of the correlation function, this simplification could be justified because the region where both amplifications of gauge fields ~$\hat{A}_+(\bm{p}), \ \hat{A}_+(\bm{k}-\bm{p})$ ~are maximal is the most dominant contribution to the integration and furthermore we can extend ~$x_1$ ~from ~$0$ ~to ~$\infty$ ~since \eqref{eq: A} has most of its support at ~$(8\xi)^{-1} \lesssim x_1 \lesssim 2\xi$.
 In this calculation, we set ~$\hat{\bm{k}} = (0, 0, 1)$ ~and ~$\bm{p} = p(\sin\theta\cos\phi, \sin\theta\sin\phi, \cos\theta)$ ~which corresponds to the following polarization vector
\begin{equation}
\bm{e}^\pm(\hat{\bm{p}}) = \dfrac{1}{\sqrt{2}}(\cos\theta\cos\phi \mp i\sin\phi, ~\cos\theta\sin\phi \pm i\cos\phi, ~-\sin\theta) \ .
\end{equation}
 Then for large ~$\xi$, we can write
\begin{equation}
f^+_h(\xi) \simeq \dfrac{4.3 \times 10^{-7}}{\xi^6} \ , \qquad f^-_h(\xi) \simeq \dfrac{9.2 \times 10^{-10}}{\xi^6} \ .
\end{equation}
 We can see that ~$f^+_h(\xi)$ ~is larger than ~$f^-_h(\xi)$ ~by a factor ~$10^{-3}$.
 This numerical discrepancy comes from the suppression factors ~$|e_{\mp}^l(\hat{\bm{k}})e^{+}_l(\hat{\bm{p}})|^2|e_{\mp}^m(\hat{\bm{k}})e^{+}_m(\hat{\bm{k}-\bm{p}})|^2$ ~in the integration.
 Therefore, the total power spectrum of tensor modes is given by
\begin{align}
\langle\hat{h}_{\pm}(\bm{k})\hat{h}_{\pm}(\bm{k}')\rangle &= \langle\hat{h}_{V\pm}(\bm{k})\hat{h}_{V\pm}(\bm{k}')\rangle + \langle\hat{h}_{S\pm}(\bm{k})\hat{h}_{S\pm}(\bm{k}')\rangle \\
&\equiv (2\pi)^3\delta^3(\bm{k} + \bm{k}')\dfrac{2\pi^2}{k^3}\Delta^\pm_h(k) \ .
\end{align}
 In the super-horizon limit, we get
\begin{align}
\Delta_h^+(k) &\simeq \dfrac{H^2}{\pi^2M_p^2}\left( 1 + \dfrac{2H^2}{M_p^2}f^+_h(\xi)e^{4\pi\xi} \right) \ , \\
\Delta_h^-(k) &\simeq \dfrac{H^2}{\pi^2M_p^2}\left( 1 + \dfrac{2H^2}{M_p^2}f^-_h(\xi)e^{4\pi\xi} \right)
\end{align}
and the following chirality parameter of tensor spectra
\begin{equation}
\chi \equiv \dfrac{\Delta_h^+(k) - \Delta_h^-(k)}{\Delta_h^+(k) + \Delta_h^-(k)} \ .
\end{equation}

\end{document}